\documentclass{astlb}
\usepackage{longtable}
\usepackage{lscape}
\usepackage{graphicx}
\usepackage{color}
\usepackage{setspace}
\usepackage{amsmath}

\usepackage{hyperref}

\definecolor{darkblue}{rgb}{0,0,0.9}

\def\ps1{\emph{Pan-STARRS1}}

\begin{document}

\journalinfo{2024}{50}{12}{1}[16]

\title{850 SRG/eROSITA X-ray sources associated with Pleiades stars}

\author{I.~M.~Khamitov\email{irek\_khamitov@hotmail.com}\address{1},
  I.~F.~Bikmaev\address{1},
   M.~R.~Gilfanov\address{2,3},
  R.~A.~Sunyaev\address{2,3},
  P.~S.~Medvedev\address{2}\\
$^1$\it{Kazan Federal University, Kazan, Russia\\}
$^2$\it{Space Research Institute of Russian Academy of Sciences, Moscow, Russia\\}
$^3$\it{Max-Planck Institute for Astrophysics, Garching, Germany}
}

\shortauthor{I.~M.~Khamitov et al.}

\shorttitle{SRG/\emph{e}ROSITA sources in Pleiades}
 
\submitted{\today}


\begin{abstract} Using data from the SRG/eROSITA  all-sky X-ray survey and the GAIA-based  catalog of 2,209 members of the Pleiades open star cluster, we found 850 X-ray sources associated with the cluster stars. Over 650 of them were detected in X-rays for the first time. At the distance of the Pleiades, the nominal sensitivity of eROSITA corresponds to a luminosity of $L_X \sim 1.6 \cdot 10^{28}$  erg/s in the 0.3--2.3 keV  band. The eROSITA sources associated with  Pleiades stars have a total luminosity of $L_{X,tot} \sim 1.3 \cdot 10^{32}$ erg/s , a million times greater than the X-ray luminosity of the quiet Sun.

Strong X-ray variability, more than 10 times, was recorded for 27 sources. Most of them are known as eruptive optical variables of the dM class.

The value of $R_X=log(L_X/L_{bol})$ increases with decreasing effective temperature of the star from $R_X\approx -5$ to $R_X\approx -2$. The distribution of stars over $R_X$ is bimodal, with the left peak at $R_X\sim-4.3$ being formed by stars of FGK classes, and the right peak at $R_X\sim-3.1$ being mainly populated by M--stars.

The relation between $R_X$ and the Rossby number $Ro$ depends on the spectral class. For K- and M- stars, at low Rossby numbers  $R_X\sim -3$ and depends weakly on $Ro$. At $Ro\ga 0.25$, a rapid drop in $R_X$ is observed for K stars, while in our sample there are no M stars with large Rossby number. Most of F- and G- stars appear to have smaller $R_X\sim -4.5$, however, our sample size is insufficient for a more detailed characterization of their $R_X-Ro$ dependence.

\end{abstract}

\section{Introduction}

The Pleiades open star cluster has 2226 known members (according to \textit{GAIA DR3}, \cite{2023AA...678A..75Z}) with ages on the order of 120-130 million years and at an average distance of 135 pc. A  detailed review on the age and distance studies of the cluster is given in \cite{2019A&A...628A..66L}. Most members of the cluster are M-class stars with a total mass on about $\sim $800$~M_{\odot}$. 
The Pleiades are perhaps one of the most well-studied cosmic ecosystems. Due to the proximity and relatively high brightness of the members of the cluster (some of them are visible even to the naked eye), the Pleiades are from an instrumental point of view a convenient target for observations. 
From a scientific point of view, the Pleiades stars, close in age and chemical composition, represent a remarkable evolutionary laboratory -- both for studying the distribution of physical characteristics of stars at an early stage of evolution and for the problems of planet formation and habitability. This explains the continuing interest in them by researchers and science missions over a wide range of the wavelength.

The discovery of X-ray emission from stars of different spectral classes, with a wide range of luminosities within the same class, by the \textit{Einstein Observatory} opened studies of stars in the X-ray band\citep{1981ApJ...245..163V}. One of the main targets of such studies has become the Pleiades region.  
Almost every X-ray observatory -- such as -- \textit{Einstein} \citep{1979ApJ...230..540G}, \textit{ROSAT} (\cite{1991Natur.349..579T}, \cite{1986SPIE..597..208P}), \textit{Chandra} \citep{2002PASP..114....1W} and \textit{XMM-Newton} \citep{2001A&A...365L...1J} , have made long-term observations of the Pleiades region. 
Depending on the depth, sensitivity, and field coverage, the information  on the X-ray sources of the Pleiades was obtained, but mostly in a small region of the cluster's core. Thus, the \textit{Einstein} Observatory conducted 14 pointings of the Pleiades region between August 13, 1979 and February 8, 1981. The duration of the pointings varied from 1.3 to 14.2 fs (\cite{1985ApJ...292..172M}, \cite{1990ApJ...348..557M}). 
With a field-of-view size of $1^o \times 1^o$, the central region of the cluster with a size of $2^o \times 2^o$ was covered more than once. In total, the \textit{Einstein} Observatory detected 85 X-ray sources of the Pleiades, 20 of which were in the central region.  
The observatory \textit{ROSAT}, carried out observations both in the all-sky survey mode \citep{1993A&A...277..114S} and in the long-duration pointing mode (\cite{1994ApJS...91..625S}, \cite{1996ApJS..102...75M}).
The \textit{ROSAT} observatory carried out an all-sky survey between July 30, 1990 and January 25, 1991.  To ensure equal sensitivity and comparison with the \textit{Einstein} data, the search for X-ray sources in the Pleiades was limited to the $2^o \times 2^o$ region of the cluster's core. 
As a result, 24 X-ray sources were found. Three regions of the Pleiades cluster were observed by the \textit{ROSAT} observatory in the long-duration pointing mode: the central, northeastern, and northwestern regions. Each region, with a field of view of radius $\sim1^o$, was exposed by the \textit{ROSAT} observatory for approximately 31, 20, and 26 ksec, respectively.
To investigate the X-ray stellar variability, the central region was observed three times: in February and August 1991 and in August 1992. The other two regions with a small overlap of the central one were observed in September 1991. A total of 171 stars in the Pleiades cluster were identified, of which 99 sources were detected in the central region.
As for the observations of the \textit{CHANDRA} observatory, the core region of the Pleiades cluster was observed twice at 38.4 and 23.6 ksec, respectively, on September 18, 1999 and March 20, 2000 (\cite{2001AJ....121..337K}, \cite{2002ApJ...578..486D}). In a field of $17'\times 17'$ among 99 X-ray sources, 23 stars belonging to the Pleiades were identified.
Long-term monitoring observations of 8 selected Pleiades stars have also been made by the \textit{XMM-Newton} Observatory, with the first observations made on September 1, 2000 for a duration of $\sim40$~ksec \citep{2003MNRAS.345..714B}. 
And in February 2015, 12 known Pleiades stars were observed simultaneously with optical observations by the \textit{Kepler} space observatory to study stellar flare activity in both the X-ray and optical bands \citep{2019A&A...622A.210G}.

The \textit{eROSITA} \citep{2021AA...647A...1P} telescope on board the \textit{SRG} orbital X-ray observatory \citep{2021AA...656A.132S}  covered the region of the sky around the Pleiades in the course of 5 all-sky surveys separated by 6 months.  
In this paper, we present the most comprehensive catalog to date of 850 X-ray sources associated with Pleiades stars.  
The identification was made using the \textit{eROSITA} X-rayu source catalog and the optical catalog of the Pleiades cluster members \citep{2023AA...678A..75Z}. The catalog of X-ray sources, together with archival data from previous missions, makes it possible to study the outburst activity of cluster stars on the time scale of up to 30--40 years, in addition to their evolutionary and population properties. 

\begin{figure*}
  \centering

\includegraphics[width=1.75\columnwidth,height=1.75\columnwidth]{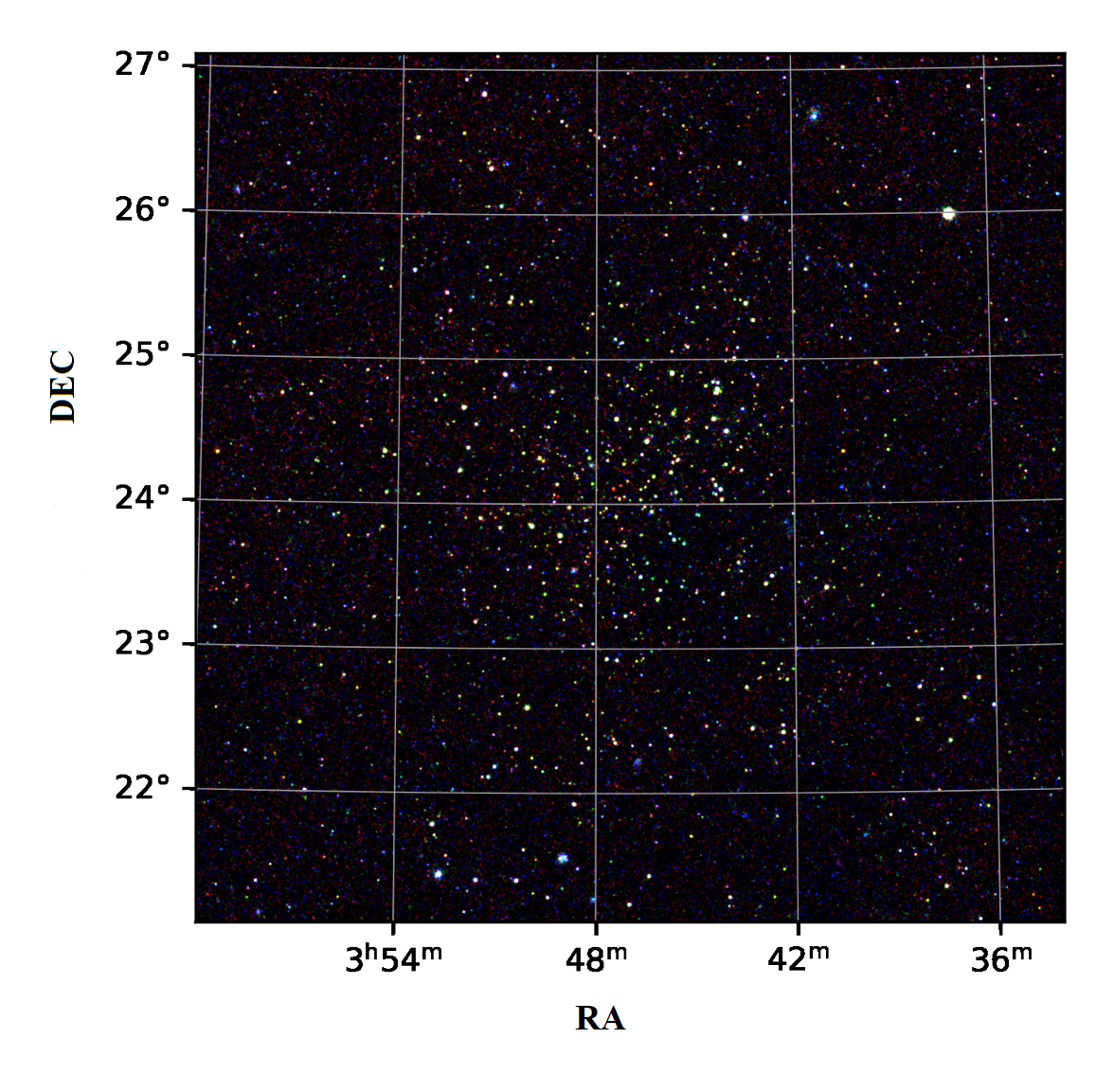}

  \caption{X-ray image of the $6^o \times 6^o$ region of the Pleiades open star cluster was formed using data of the \textit{eROSITA} X-ray telescope of the SRG space observatory. The RGB image was plotted using the following energy ranges: from 300 to 600 eV in red, from 600 eV to 1.0 keV in green, and from 1 to 2.3 keV in blue. }
  \label{fig:Pleiades_eROSITA}
  
\end{figure*}

\section{Identification of X-ray sources in the Pleiades}

\renewcommand{\arraystretch}{1.21}
\renewcommand{\tabcolsep}{3pt}
\begin{table*}
\small
\caption{The \textit{eROSITA} X-ray sources associated with stars in the Pleiades. 
\label{cat_src}}
\begin{centering}
\begin{tabular}{lccccrrrrccrl}
\hline
\hline
SRGe+ & Gaia DR3+  & RA         & DEC        & G      & $r98$  & sep    & N & $L_X$     & $L_X\_err$  & \textit{\footnotesize{log}}$\frac{L_X}{L_{bol}}$ & $X_{var}$ & $SpT$ \\
      &            & ($^\circ$) & ($^\circ$) & (mag)  &($''$)  & ($''$) &   &  &   &  (dex)                &           &       \\

\hline
J025845.9+212815 & 108606322618018432 & 44.6915 & 21.4702 & 9.3 & 3.0 & 2.1 & 1 & 9.25 & 0.82 & -4.01 & 1.5 & F7V  \\
J030319.1+314527 & 135368508653824768 & 45.8281 & 31.7582 & 15.0 & 8.6 & 5.1 & 1 & 0.60 & 0.14 & -3.16 & $>$1.8 & M2.5V  \\
J030417.5+190747 & 59777943236934528 & 46.0721 & 19.1281 & 15.7 & 10.8 & 6.3 & 1 & 0.51 & 0.16 & -3.15 & $>$1.3 & M3V  \\
J030426.4+331703 & 136166066901394048 & 46.1100 & 33.2838 & 16.5 & 6.3 & 0.9 & 2 & 0.85 & 0.18 & -2.64 & $>$9.1 & M3.5V  \\
J030649.7+174135 & 34825351478452352 & 46.7092 & 17.6929 & 16.5 & 8.6 & 7.2 & 1 & 0.46 & 0.14 & -2.91 & $>$1.0 & M3.5V  \\
J030801.4+291427 & 122767864881453568 & 47.0055 & 29.2418 & 16.1 & 6.3 & 4.1 & 1 & 0.57 & 0.12 & -2.82 & $>$2.0 & M3.5V  \\
J030805.2+284213 & 121962197736025216 & 47.0235 & 28.7039 & 14.6 & 5.6 & 5.8 & 1 & 1.21 & 0.17 & -2.93 & $>$7.2 & M2.5V  \\
J030936.2+245422 & 111721552594418304 & 47.4008 & 24.9065 & 12.1 & 4.6 & 1.3 & 2 & 2.46 & 0.26 & -3.48 & 10.5 & K5V  \\
\hline\hline
\end{tabular}
\end{centering}
\\
The complete table is available in the archive of the Astronomical Data Center in Strasbourg.
The columns have the following designations: 1) identification number in the \textit{SRG/eROSITA} catalog; 2-5) identification number, Right Ascension and Declination in degrees (J2000.0), stellar magnitude in the G band according to Gaia DR3; 6)  \textit{r98}- radius 98\% of the position error circle of the \textit{eROSITA} sources in arcsec; 7) distance between the X-ray and optical source positions in arcsec; 8) $N$ - number of optical sources \textit{Gaia DR3} inside $r98$; 9) X-ray luminosity of the source estimated on the base of 5 surveys in the range 0.3---2.3 keV in units of $ \times 10^{29}$~erg/s; 10) uncertainty on the X-ray luminosity of the source in units of $ \times 10^{29}$~erg/s; 11) logarithm of the ratio of the X-ray luminosity $L_X$ based on  5 \textit{eROSITA} surveys   in the range 0.3-2.3 keV to the bolometric luminosity $L_{bol}$ of the star; 12) $X_{var}$ -- X-ray variability factor, defined as the ratio between the maximum and minimum flux values recorded in 5 \textit{eROSITA}  surveys taking into account the flux measurement error; 13) $SpT$ -- spectral class of the star.
\end{table*}

\subsection{eROSITA telescope data}

The Pleiades region was observed by the \textit{SRG} observatory with relatively uniform  coverage over the field with an exposure of $\sim750$~sec. As a result, a nominal flux limit of $F_X\sim7.4\times 10^{-15}~erg/s/cm^2$ was achieved in the energy range of 0.3--2.3 keV.   Figure \ref{fig:Pleiades_eROSITA} shows an X-ray image in pseudo-colors of the $6^o\times6^o$ Pleiades region. The RGB image is compiled from \textit{eROSITA} data in three energy ranges: from 300 to 600 eV (red), from 600 eV to 1.0 keV (green), and from 1 to 2.3 keV (blue).  To identify X-ray sources in the Pleiades, we used the \textit{eROSITA} X-ray point source catalog of the summed data of 5 sky surveys in the 0.3---2.3 keV range.  The X-ray fluxes given in the \textit{eROSITA}  catalog were recalculated for the spectrum of thermal emission  of  optically thin plasma with temperature of 150 eV assuming a zero column of neutral hydrogen in the line of sight, $NH=0$, for the same 0.3--2.3 keV energy range. For a temperature of 300 (500) eV, the presented luminosities should be multiplied by a factor of 0.83 (0.81). Moderate values of column density, $NH\sim 10^{20} cm^{-2}$ change these values by a few percent. We did not attempt to apply the bolometric correction of X-ray luminosities because of its large uncertainty. The catalog was filtered by the detection confidence threshold corresponding to $4\sigma$ (threshold on the likelihood value $\geq10$). In addition, we filtered out sources for which the radius of the 98\% positional uncertainty exceeded $20''$

\subsection{Identification of optical components.}

As a reference catalog for the identification of \textit{eROSITA}  X-ray sourcesin the Pleiades, we used the most complete and comprehensive catalog of members of the nearest open clusters, including the Pleiades, presented in \cite{2023AA...678A..75Z}. The selection of cluster members in this catalog is based on the exceptional astrometry of the \textit{GAIA DR3} catalogs in the five- or six-dimensional space of spatial and kinematic parameters. Hereafter, we will refer to it as the ''6d-catalog''. The difference between the choice of spatial and kinematic parameters in five-dimensional or six-dimensional space is related to the presence or absence of data on the radial velocities of stars. Within the tidal radius of the cluster ($r_{tidal}=11.28 \pm 0.03$~pc), 1355 so-called bonafide members were identified.   The 6d-catalog of the Pleiades includes members up to 3 tidal radii away from the center of the cluster, 2226 in total.  Optically brightest stars  with $G<5^m$ were excluded from further analysis.  This constraint is due to the  feature of the \textit{eROSITA} detectors associated with leakage in the ultraviolet part of the spectrum. In addition, since the Russian \textit{eROSITA} consortium is responsible for processing data for the eastern hemisphere of the Galaxy, the Pleiades stars only in this region of the sky were considered. Note that this constraint only affects  the peripheral part of the Pleiades, which contains about $\sim 1\%$ of the stars in the cluster, and does affect the region inside the tidal radius, the study of which is the focus of this paper. All in all, 2209 stars from the 6d-catalog were used for further analysis. 

The \textit{eROSITA} catalog of X-ray sources was cross-correlated with the 6d Pleiades catalog using the match radius equal to $r98$ radii of X-ray sources. 
A total of 850 \textit{eROSITA} X-ray sources were identified that contained at least one Pleiades star in their error circle, of which  38 sources contain two Pleiades stars in their error circle.

\section{Catalog of X-ray bright stars in the Pleiades based on \textit{{\footnotesize e}ROSITA} data.}

Part of the catalog of 850 X-ray \textit{eROSITA}  sources associated with stars in the Pleiades is presented in Table \ref{cat_src}. The complete catalog will be available electronically at the Astronomical Data Center in Strasbourg.

\begin{figure*}
  \centering
  \includegraphics[width=1.8\columnwidth,height=0.9\columnwidth]{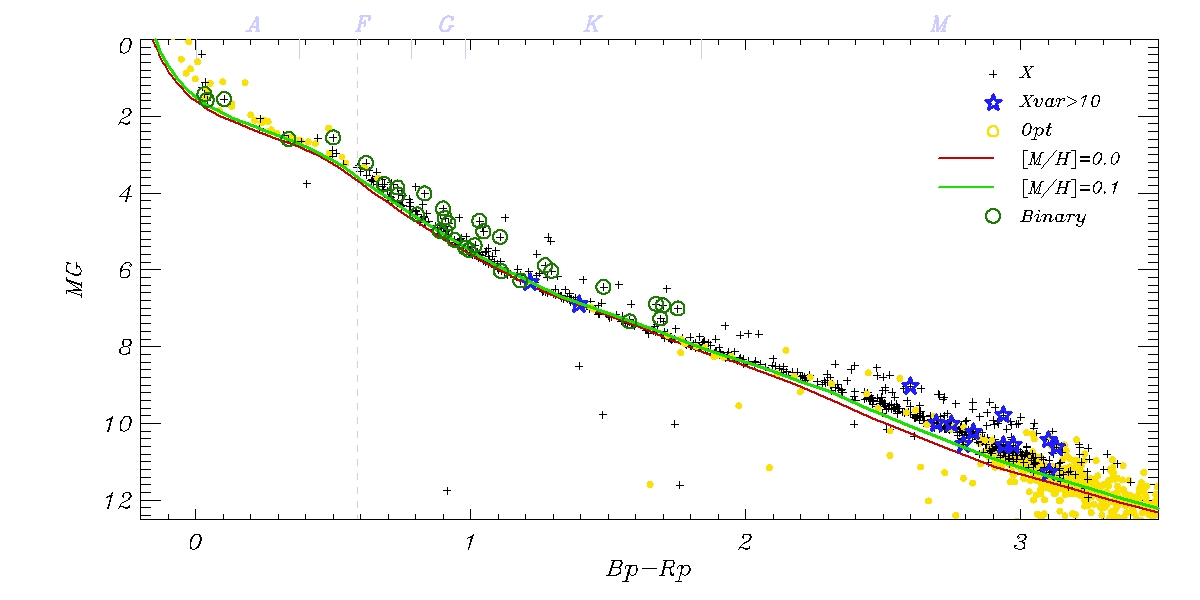}

  \hspace{1pt}
  \caption{Color (Bp-Rp) --- absolute stellar magnitude (MG) dependence of Pleiades stars within the tidal radius.  Stars with X-ray emission detected by \textit{eROSITA} are shown with black crosses, and yellow circles show stars without X-ray match. X-ray detected stars which are  spectroscopically confirmed binaries \citep{2021ApJ...921..117T} are marked with green circles. The isochrones calculated using the \textit{PARSEC v1.2S} \citep{2012MNRAS.427..127B} resource, with a fixed stellar age of 125 million years and metallicities: $[M/H]=0.0$ (solar) and $[M/H]=0.1$, are shown in red and green lines respectively.  The approximate boundary of Main Sequence stars with the presence of a convective zone is shown by a vertical dashed line.   X-ray sources with strong variability $X_{var}>10$ are marked with blue asterisks.}
  \label{fig:BpRp_MG_dtidal}
  
\end{figure*}

The bolometric luminosities and spectral classes were estimated by interpolating the dependence of these parameters on the absolute stellar magnitude (\textit{MG}) in the G band (\textit{GAIA DR3}) for main-sequence stars.  The version of Mamajek's tables ($Version~2022.04.16$) originally published in \cite{2013ApJS..208....9P} was used for this purpose.  An estimate of the effective temperature ($T_{eff}$) and the mass in solar mass units ($M/M_\odot$) was obtained in the same way.

In determining $L_{bol}$, $SpT$, $T_{eff}$, and $M/M_\odot$, the binaries were not taken into account. According to \citep{2021ApJ...921..117T}, only 35 spectroscopic binary objects have been identified in our catalog. However, the actual number of binary stars in the Pleiades must be much larger. In particular, the single-star assumption can lead to incorrect Gaia solutions for binary stars and make it difficult to classify them as cluster members. In general, we note that for systems with q close to 1, the parameter $R_X$ will not be strongly affected (the X-ray luminosity will increase about the same as the bolometric luminosity of such a system). In contrast, systems with small values of q at comparable absolute values of the X-ray luminosities of stars of early and late spectral classes (Fig.\ref{fig:Teff_LX}) can lead to an underestimation of the parameter $R_X$ - approximately proportional to the ratio of the bolometric luminosities of the binary system components.  In the color-luminosity diagram, in the limiting case of spectroscopic binary systems with $q\sim 1$, the recorded absolute stellar magnitude (MG) will be shifted by 0.75 stellar magnitude. In Fig.\ref{fig:BpRp_MG_dtidal} just above the main sequence, a small group of such photometric double systems is clearly visible.  

In Table \ref{cat_src}, X-ray sources with more than one Pleiades star within $r98$ are marked with an (*) in the identification number of the \textit{ eROSITA} source. 
The angular resolution of the \textit{eROSITA} telescopes is not sufficient to unambiguously identify these X-ray sources.

The X-ray variability factor $X_{var}$ was defined as the ratio between the maximum and minimum fluxes recorded during the 5 sky surveys. Fluxes in individual surveys were measured by forced PSF- photometry.  For this purpose, the distribution of counts in the X-ray image was approximated by the telescope point source response function, taking into account the background map. The approximation was performed by the maximum likelihood method considering a Poisson distribution of samples, and the source position was fixed at the position determined from the sum of data from all 5 surveys, i.e., the only approximation parameter was the flux from the source.  In choosing the maximum and minimum flux values, in those surveys in which the source was recorded with a likelihood value of at least 6 (corresponding to a confidence level of $\approx 3\sigma$ for a Gaussian distribution), the measured flux value was used. In the case where a survey had a source detection confidence of less than 6, the $3\sigma$ upper limit of the flux in that survey was used in determining the minimum flux and that measurement was not used in determining the maximum flux. In case no source was detected in any of the sky surveys with a detection likelihood $>$6, the value $X_{var}$ for that source was considered as undetermined.  

\section{Properties of stars with X-ray emission within the tidal radius of the Pleiades cluster.}

To investigate the properties of X-ray stars in the Pleiades, we restricted ourselves to the region within the tidal radius (11.2 pc). This sample includes 1349 stars from the 6d-catalog. The number of \textit{eROSITA} sources associated with Pleiades stars is 688, i.e. \textit{eROSITA} detects X-ray emission from more than half of the stars within the Pleiades tidal radius.

\subsection{Color-luminosity diagram.}

The color-luminosity diagram for stars within the tidal radius of the Pleiades is shown in Fig. \ref{fig:BpRp_MG_dtidal}. Stars with X-ray emission are shown with black crosses on it, and stars not detectable with \textit{eROSITA} are shown with yellow circles. X-ray sources with multiplicity detected based on spectral observations \citep{2021ApJ...921..117T} are marked with green circles. Also, stars with X-ray variability factor $X_{var}>10$, i.e., those sources that show X-ray flux variability of more than an order of magnitude, are shown with blue symbols. It should be noted that the bulk of the strongly variable X-ray sources belong to M-class dwarfs.

In Fig. \ref{fig:BpRp_MG_dtidal}, an isochron calculated from the \textit{PARSEC v1.2S} \citep{2012MNRAS.427..127B} resource, with a fixed stellar age of 125 million years and solar metallicity, is shown with a red line. The metallicity of stars in the Pleiades was determined using 10 members of the cluster based on high spectral resolution spectroscopy and photometric methods (\cite{2016A&A...585A.150N} and references therein). The obtained estimates $[M/H]_{sp}=-0.01\pm0.05~dex$ and $[M/H]_{phot}=-0.04\pm0.11~dex$ are close to the solar metallicity.   In Fig. \ref{fig:BpRp_MG_dtidal} the systematic displacement of M-class main-sequence (MS) stars from this model isochrone is observed. This behavior is true for both X-ray sources and Pleiades stars with X-ray emission levels below the \textit{eROSITA} sensitivity limit. The difference is slightly smaller for the isochrone with metaicity $[M/H]=0.1$ (green line in Fig. \ref{fig:BpRp_MG_dtidal}).   This displacement may be due to flaws in the model isochrone used or more complex effects (see, for instance, \cite{2012MNRAS.424.3178B}), but investigating its causes is beyond the scope of this study.

In Fig. \ref{fig:BpRp_MG_dtidal}, a group of sources located above the observed MS, the so-called photometric double systems, is also clearly visible. In addition to the multiplicity of systems, the presence of spots in active stars can lead to reddening of the measured colors and have a smaller effect on the integrated flux in the visible region. Calculations performed for solar metallicity in Pleiades stars of late spectral class show that the colors of fast rotators are well described by models with high spot filling factor. In contrast, slowly rotating sources are described by spotless  isochrones \citep{2020ApJ...891...29S}.

Stars with X-ray emission are detected in all spectral ranges, from early A-classes to late M-dwarfs (M5V), but their relative fraction increases later than the F5 class, from which the conditional boundary of the convective zone begins.  This boundary is shown as a vertical dashed line in Fig. \ref{fig:BpRp_MG_dtidal}. The absence of X-ray stars at the right end of the main sequence, among stars fainter than 11 absolute magnitude, is explained by the sensitivity limit of \textit{eROSITA}.

\begin{figure}
  \centering
  \includegraphics[width=\columnwidth]{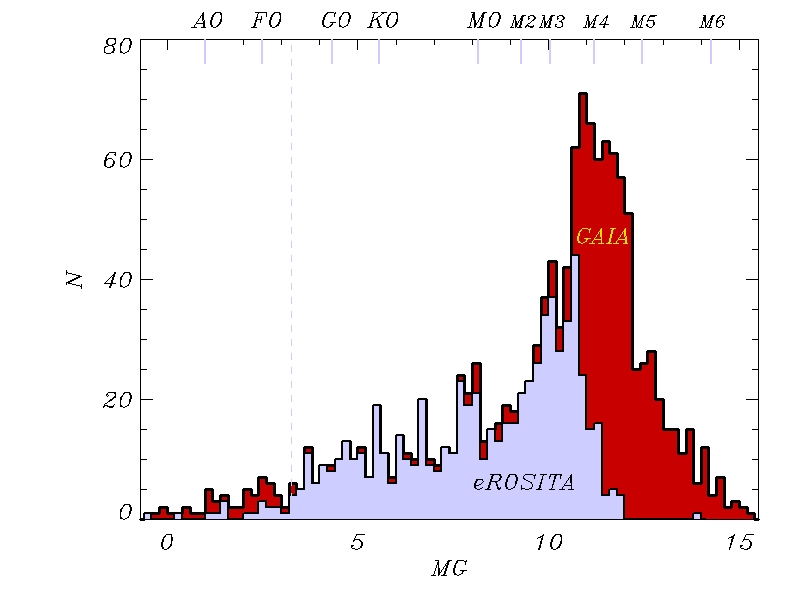}
        
  \hspace{1pt}
  \caption{Distribution of Pleiades stars over absolute stellar magnitude (MG). The distribution of all stars within the tidal radius of the cluster is shown in red; stars with X-ray counterparts  are shown by gray histogram. The vertical dashed line marks the approximate boundary of stars with the presence of a convective zone.}
  \label{fig:histogramm_dtidal}
  
\end{figure}

\begin{figure}
  \centering
  \includegraphics[width=\columnwidth]{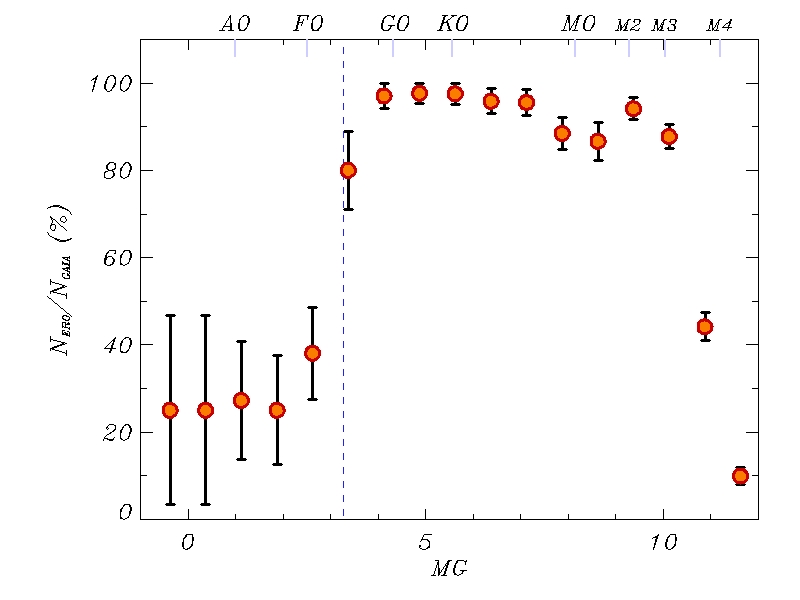}
        
  \hspace{1pt}
  \caption{The fraction of stars within the tidal radius of the Pleiades detected in  X-ray band  as a function of absolute stellar magnitude (MG).  The vertical dashed line shows the approximate boundary of stars with the presence of a convective zone.  }
  \label{fig:fraction_dtidal}
  
\end{figure}

\begin{table}
\small
\caption{ The fraction of Pleiades stars with X-ray emission as a function of the spectral type. Only stars within the tidal radius are taken into account.} 
  \label{tab:Nstype}
  \vskip 2mm
  \renewcommand{\arraystretch}{1.05}
  \renewcommand{\tabcolsep}{0.05cm}
  \centering

\begin{tabular}{ccccc}
       \noalign{\vskip 3pt\hrule\vskip 5pt}
Spectral  &  $N_{opt}$  & $N_X$      &  $N_X/N_{opt}$  & $\sigma$   \\
range      &  (GAIA)     & (\textit{eROSITA})  &  (\%)           &  (\%)  \\
\hline

earlier F5   &  57   &   19   &   33.3  & 6.2 \\
F5-G0      &  41   &   39   &   95.1  & 3.4 \\
G0-K0      &  64   &   62   &   96.9  & 2.2 \\
K0-M0      & 192   &  180   &   93.7  & 1.7 \\
M0-M1      &  51   &   43   &   84.3  & 5.1 \\
M1-M2      &  38   &   34   &   89.5  & 5.0 \\
M2-M3      & 118   &  110   &   93.2  & 2.3 \\
M3-M4      & 309   &  173   &   56.0  & 2.8 \\
M4-M5      & 321   &   27   &    8.4  & 1.5 \\
later M5 & 158   &    1   &    0.6  & 0.6 \\
\hline
total      & 1349  &  688   &   51.0  & 1.4 \\   

  \end{tabular}
\end{table}

\subsection{Distribution of X-ray emitting stars in the Pleiades by spectral classes.}

The distribution of stars within the tidal radius of the Pleiades by spectral class is shown in Fig. \ref{fig:histogramm_dtidal}. The distributions of all stars and stars with X-ray are shown in the figure. The histograms are plotted by absolute stellar magnitude with binning by $0.^m75$. As in Fig. \ref{fig:BpRp_MG_dtidal}, the conditional boundary of MS stars with the presence of a convective zone is shown by the vertical dashed line.

The dependence of the fraction of stars with X-ray emission on the spectral class is shown in Fig. \ref{fig:fraction_dtidal}.  The errors in each bin were calculated assuming a binomial distribution of the number of X-ray sources \footnote{ If $N_{gaia}$ is the total number of sources in a bin from which X-ray emission is recorded from $N_{ero}$, then $p=N_{ero}/N_{gaia}$ -- an estimation of the fraction of stars with X-ray emission in a given bin.  Assuming a binomial distribution, the expected variance $Var(N_{ero})=N_{gaia}*p*(1-p)=N_{ero}*(1-N_{ero}/N_{gaia})$. Consequently, $\sigma=\sqrt{N_{ero}*(1-N_{ero}/N_{gaia})}$}. The results of the calculations of the fraction of stars with X-ray in broad intervals by spectral class are summarized in Table \ref{tab:Nstype}.

In terms of the presence of detectable X-ray emission, three intervals of spectral classes can be distinguished in Figs.  \ref{fig:histogramm_dtidal} and \ref{fig:fraction_dtidal} : 1) earlier F5-class stars, 2) F5-M3-class stars, and 3) later M3-class stars.  The first group is characterized by the absence of a convective zone in the stars and low X-ray luminosity, despite the large luminosity in the optical range. Accordingly, the number of stars detected with \textit{eROSITA} is small and limited by the sensitivity of the X-ray survey.  The fraction of stars with X-ray emission is only a third of the total number of \textit{GAIA} stars in this group. Taking into account that the X-ray luminosity of A-class stars (\cite{2014ApJ...786..136D}, \cite{2022AJ....164....8G}) is significantly lower than the luminosity limit reached in the \textit{eROSITA} survey at the distance of the Pleiades, the detected X-ray emission of stars in this group is most probably due to a secondary component in the binary system. The second interval in spectral classes demonstrates a high fraction of stars with X-ray emission -- above $\sim 80-90\%$. In the third spectral class interval, the absolute value of stellar luminosity in the optical and X-ray ranges decreases, and the number of stars detected by \textit{eROSITA} is limited by the sensitivity of the X-ray survey.  
 
In Fig.\ref{fig:fraction_dtidal}, a peculiarity in the behavior of the fraction of stars with X-ray emission in the spectral classes M0-M3 draws attention. This feature can be interpreted as a dip in the M0-M1 spectral classes or as a rise in the M2-M3 region. The statistical significance of this feature in our data does not exceed $\sim 2\sigma$, so we will postpone a more detailed discussion of it for future work if its existence will be confirmed with higher confidence.  

\begin{figure*}
  \centering
  \includegraphics[width=\columnwidth]{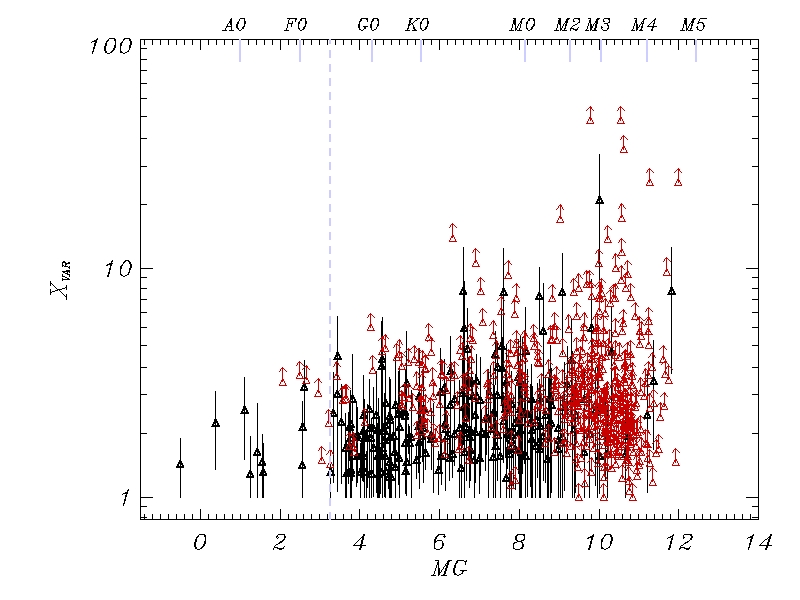}
   \includegraphics[width=\columnwidth]{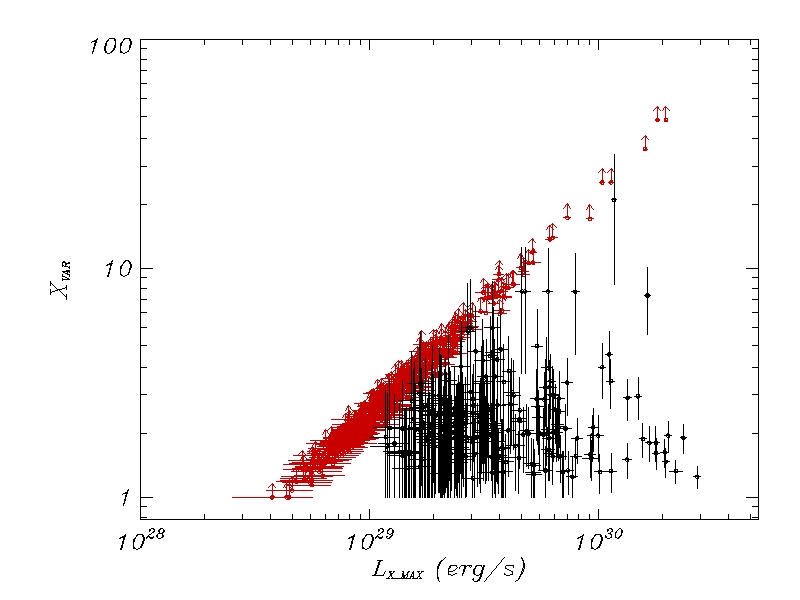}
  \hspace{1pt}
  \caption{Dependence of the X-ray variability parameter $X_{var}$ on absolute stellar magnitude (left panel) and peak X-ray luminosity (right panel). With red arrows, the lower limits on $X_{var}$ are shown for sources that are not detected in the minimum-flux state. For such sources, the lower limit on $X_{var}$ is roughly proportional to the X-ray flux, as seen in the right figure. In the left figure, the vertical dashed line marks the conventional boundary of stars with the presence of a convective zone.}
  \label{fig:xvar}
  
\end{figure*}

\subsection{X-ray variability.}

The dependence of the X-ray variability parameter $X_{var}$ introduced above on absolute stellar magnitude and peak X-ray luminosity is shown in Fig.\ref{fig:xvar}. Sources for which only a lower limit of the $X_{var}$ parameter can be determined are shown as red arrows.  It can be seen that there is an increase in X-ray variability later than the M2 class.

Variability with $X_{var} \geq 10$ is detected for 27 Pleiades sources, of which 13 are located within the tidal radius of the cluster. The 3 strongly variable sources in the $r98$ error circle contain two Pleiades stars, they are SRGeJ034422.1+244606, SRGe034707.1+234252, and 034916.7+240059.

As can be seen in Fig.\ref{fig:xvar} almost all strongly variable objects were not detected in the state with minimum flux.  They belong to M-class cool stars and are known as eruptive optical variable stars, and their strong X-ray variability is mainly associated with stellar flares. A separate paper will be devoted to their study.

Also from Fig.\ref{fig:xvar} (right panel) it can be seen that for sources with peak X-ray luminosity below $\sim 10^{29}$ erg/s there are only lower limits on the variability factor $X_{var}$. This is because the detection threshold in an individual scan is about 3 times worse than the sensitivity threshold over all data and can exceed $\sim 5\times10^{28}$~erg/s. Due to some variability and/or Poisson variations, for sources fainter than $\sim 10^{29}$~erg/s in at least one sky survey the source is not detected and according to the definition of $X_{var}$ for such sources we only obtain a lower limit on this parameter.

\subsection{Dependence of the X-ray luminosity of Pleiades stars on the effective temperature.}

\begin{figure*}
  \centering

  \includegraphics[width=1.75\columnwidth,height=0.8\columnwidth]{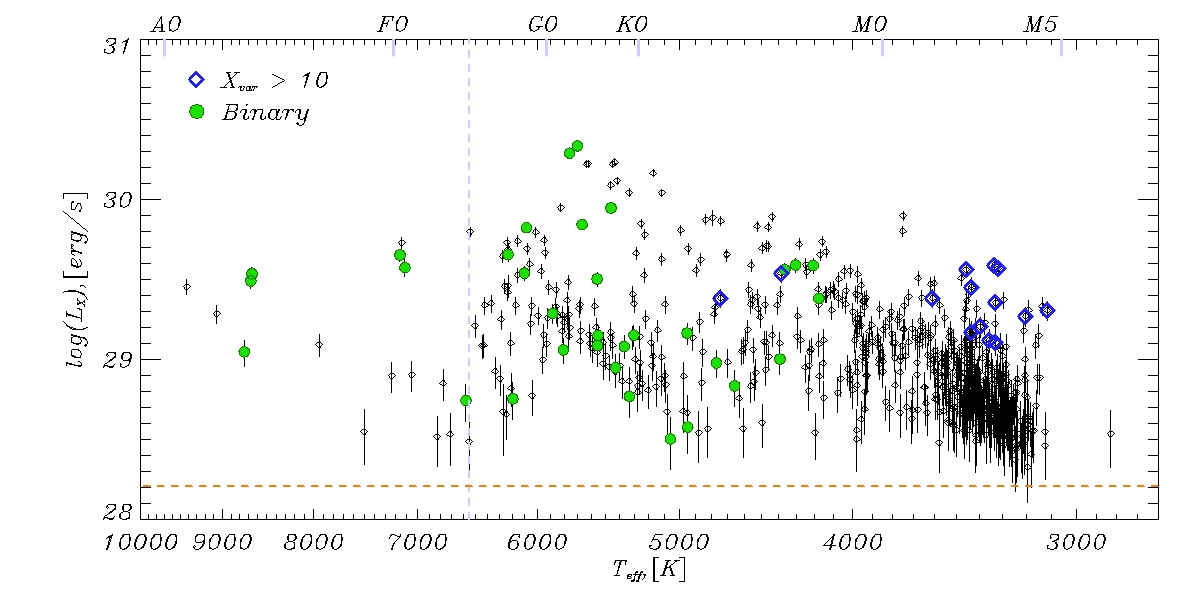}

\includegraphics[width=1.75\columnwidth,height=0.8\columnwidth]{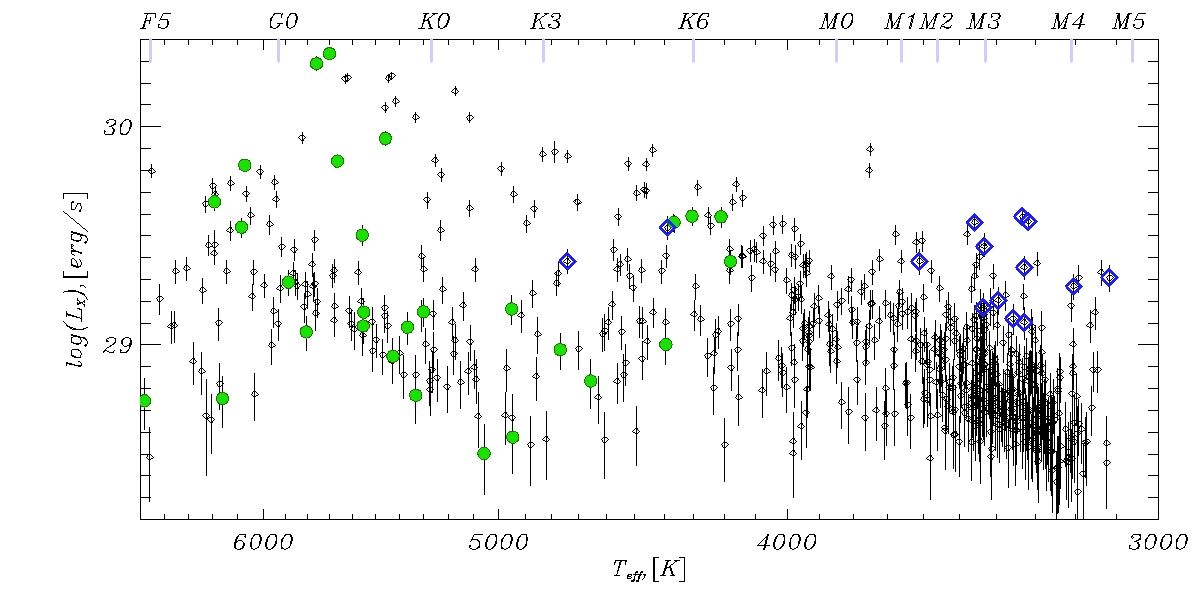}

\includegraphics[width=1.75\columnwidth,height=0.8\columnwidth]{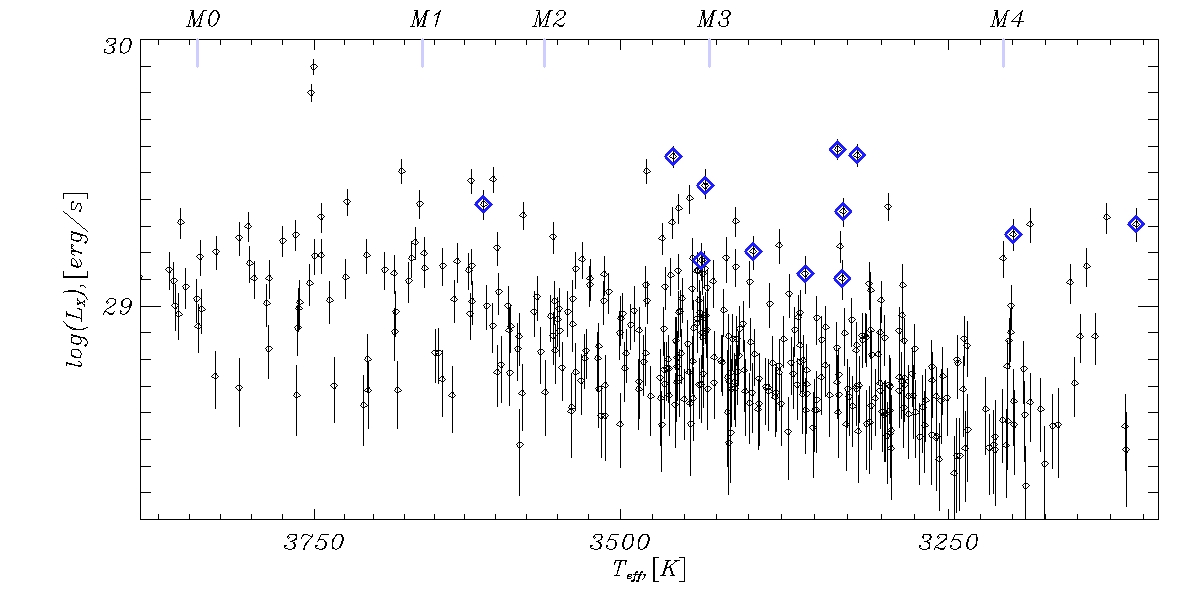}

  \caption{Dependence of the 0.3-2.3~keV X-ray luminosity  on the effective temperature $T_{eff}$ of the optical component. The strongly X-ray variable stars  are shown by blue rhombuses. The spectroscopically identified binary stars  \citep{2021ApJ...921..117T} are shown with green circles.  Top panel shows the overall view, with the red horizontal dashed line showing the nominal sensitivity of \textit{eROSITA} data at the average Pleiades distance of 135 pc. The middle panel shows the dependence for stars with the presence of a convective zone. Lower panel presents the expanded view of stars of late spectral classes.
  }  
  \label{fig:Teff_LX}
  
\end{figure*}

In the distribution of the X-ray luminosity of Pleiades stars, in general, there is no definite dependence on the effective stellar temperature. A broad range of stellar spectral classes from early (no convective zone) to late M-classes is present (Fig. \ref{fig:Teff_LX}). The maximum X-ray luminosity values are registered for solar-type stars $L_X\sim 2\times 10^{30}$~erg/s, which is two orders of magnitude higher than the \textit{eROSITA}  limiting sensitivity at an average Pleiades distance of 135 ps. It is also remarkable that in G-class stars ($5000K < T_{eff} < 6000K$) the sources with minimal X-ray luminosity significantly exceed the \textit{eROSITA} sensitivity threshold.

\subsection{The ratio of X-ray and bolometric luminosities.}

\begin{figure}
  \centering
  \includegraphics[width=1.0\columnwidth,height=0.7\columnwidth]{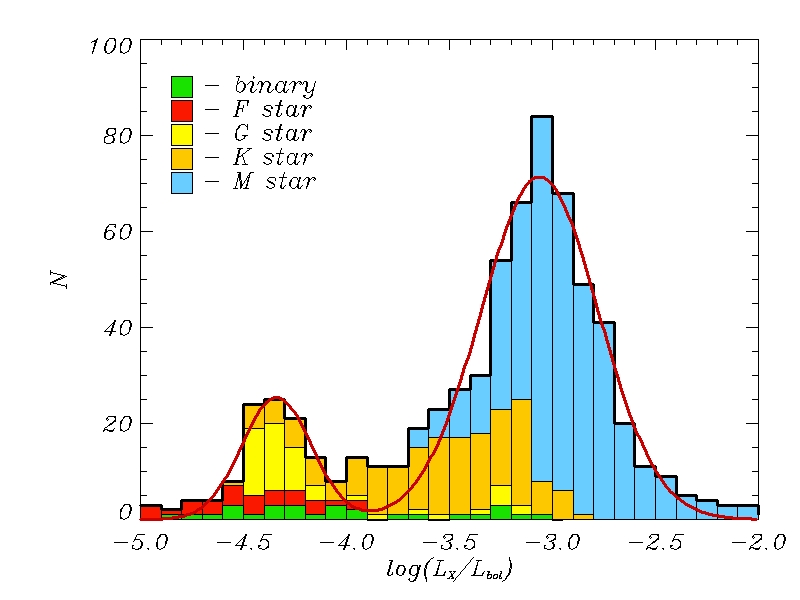}

  \hspace{1pt}
  \caption{Distribution of $R_X=log(L_X/L_{bol})$ for  Pleiades stars with \textit{eROSITA} X-ray counterparts. The red line shows the fit with two Gaussians. The peak with a smaller value of $R_X$ is formed by FGK-stars, while the second peak is populated mainly by K- and M- stars. The distribution is shown as a so-called histogram with accumulation, where the total value in a given bin on $R_X$ is defined as the sum of all histograms of different colors contributing to the bin.} 
  \label{fig:LxLbol_hist}
  
\end{figure}

\begin{figure*}
  \centering
  \includegraphics[width=1.75\columnwidth,height=0.8\columnwidth]{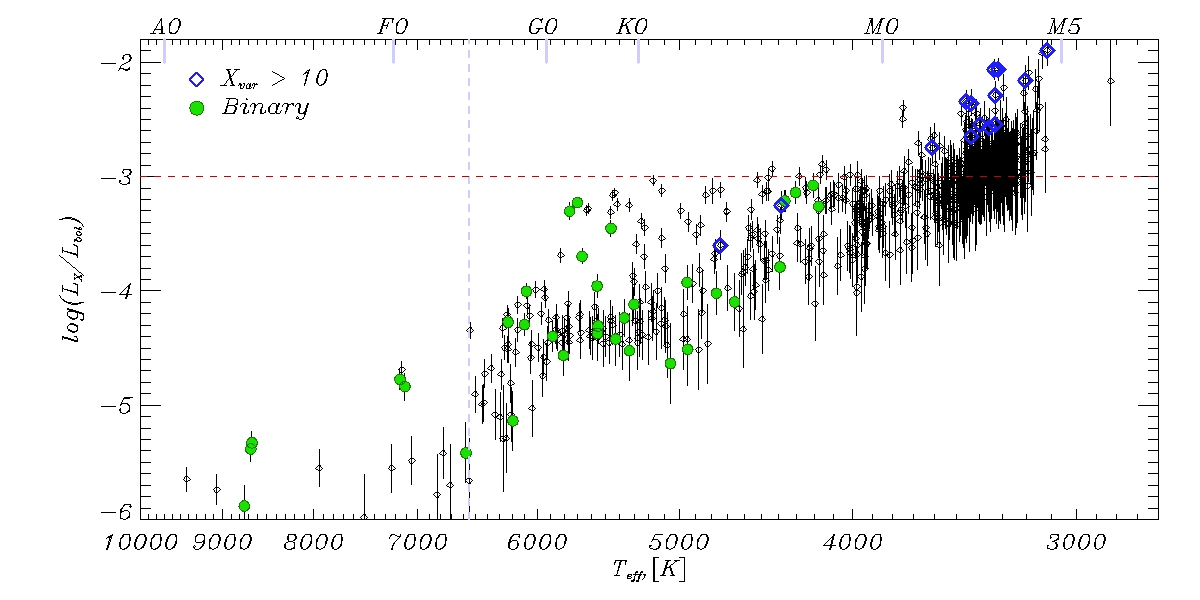}
  \includegraphics[width=1.75\columnwidth,height=0.8\columnwidth]{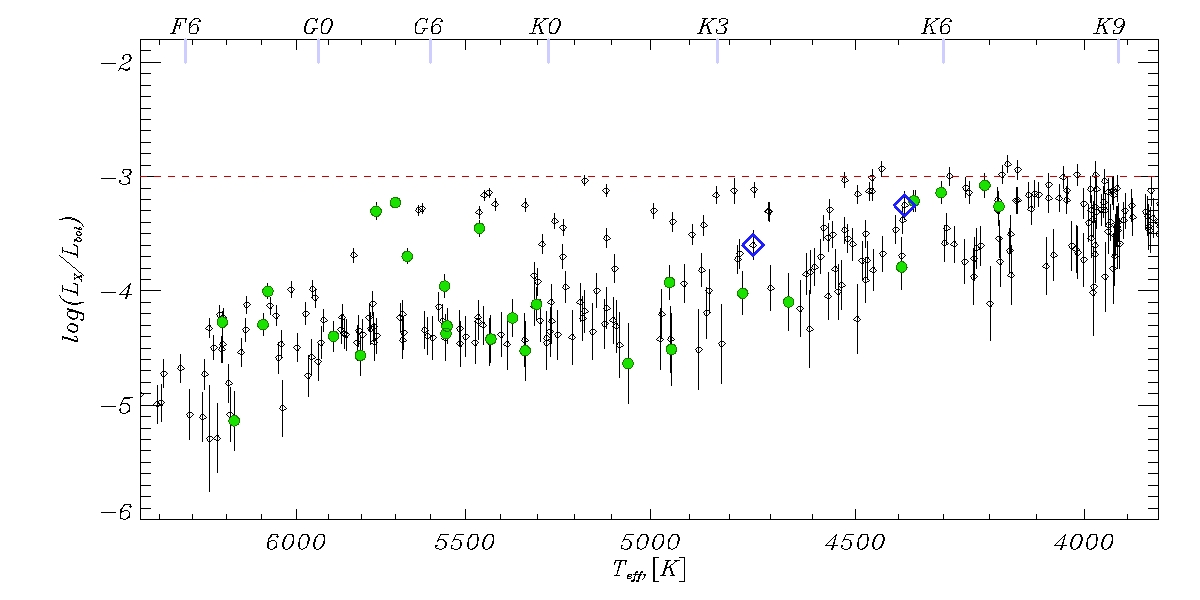}
  \includegraphics[width=1.75\columnwidth,height=0.8\columnwidth]{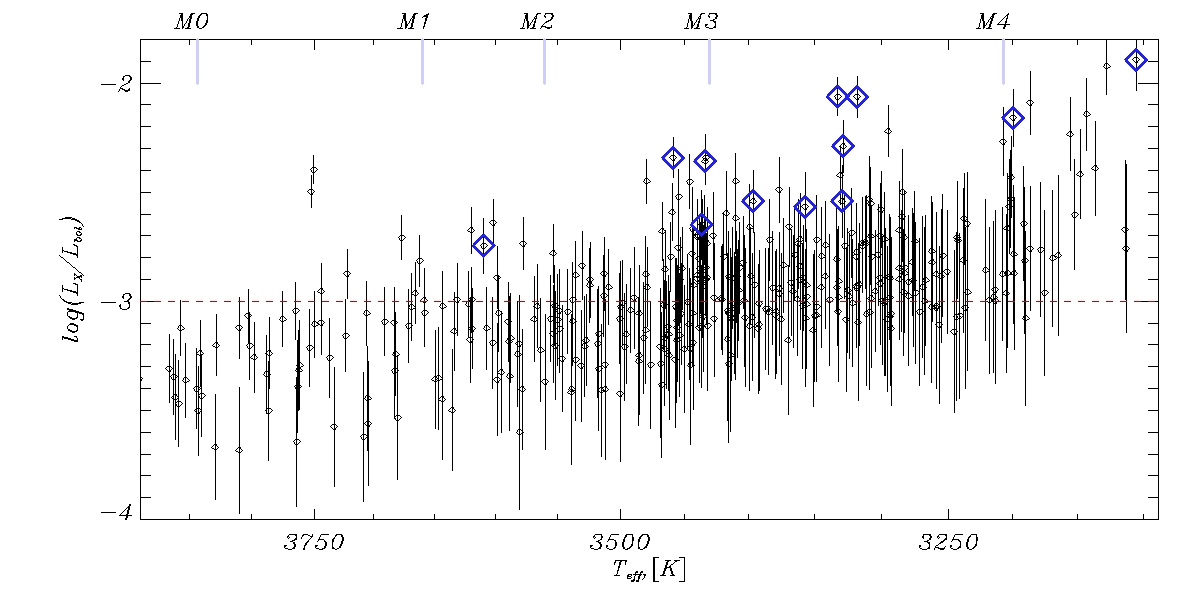}

  \caption{Dependence of the ratio $log(L_X/L_{bol})$ on the effective temperature $T_{eff}$ of the optical component. The vertical dashed line shows the conventional boundary between stars with and without convective envelope. The horizontal red dashed line corresponds to the fraction of the X-ray luminosity in $10^{-3}$ of the bolometric luminosity. Stars with X-ray variability greater than 10 are marked in blue. Green circles show spectroscopical binary stars \citep{2021ApJ...921..117T}. The top, middle and bottom panels are same as in Fig.\ref{fig:Teff_LX}. }
  \label{fig:Teff_LxLbol}
  
\end{figure*}

It is known that in stars of late solar-type spectral classes convection together with rotation leads to the generation of a magnetic dynamo at the base of the convection zone. The results of the magnetic dynamo manifest themselves as magnetic phenomena in and above stellar photospheres in the form of magnetic spots, magnetically confined coronal plasma in which periodic flares occur, and so on. It is expected that, due to the rotationally induced internal dynamo, the level of X-ray luminosity correlates with the rotational velocity of the star. However, at the level of the so-called canonical limit, $L_X/L_{bol} \approx 10^{-3}$ (\cite{1984A&A...133..117V}; \cite{ 2011ApJ...743...48W}; \cite{2024A&A...684A.121F}), saturation has taken place and the X-ray luminosity does not change anymore with increasing rotational velocity. The reason for this saturation of the X-ray emission has not yet been understood. It may be a manifestation of the physical saturation of the dynamo mechanism or a complete coverage of the stellar surface by active regions that give the largest contribution to the X-ray emission (see, for example, the review of \citealt{2009A&ARv..17..309G}).

The distribution of the value $R_X=log(L_X/L_{bol})$ is bimodal (Fig. \ref{fig:LxLbol_hist}). The peak with a smaller value of $R_X$ is formed by stars of FGK classes, while the second peak is mainly populated by M-stars. Gaussian fitting yields peak values of about -4.3 $dex$ and -3.1 $dex$, with FMHMs of 0.2 $dex$ and 0.3 $dex$, respectively.

Similar bimodality is present in other star clusters, particularly in the older-than-Pleiades clusters the $\sim$700 million year old of the Hyades and Praesepe \citep{2022ApJ...931...45N}.  The bimodal distribution of $R_X$ with maxima around -3.1 and -4.3 dex is also found for stars of the field \citep{2024A&A...684A.121F}.

The incorrect determination of $R_X$ in spectral-binary systems may be one explanation for the observed bimodality. The bolometric luminosity estimation refers to the massive component, while the X-ray emission is produced mainly by a secondary, less massive component with significant coronal activity. This may lead to the appearance of a second peak at smaller values of $R_X$.

It is also possible that the observed bimodality has a physical nature. It is interesting to note that, if so, the difference in the characteristics of coronal activity is apparent already in the first 100 million years of the evolution of main-sequence stars.  

The dependence of $R_X$ on the effective temperature of the star is shown in Fig.\ref{fig:Teff_LxLbol}.  It is obvious from the figure that $R_X$ increases with decreasing stellar effective temperature -- the $L_X/L_{bol}$ ratio increases by more than 3 orders of magnitude from stars of early spectral classes to late spectral classes. About $\sim27\%$ of the sources are characterized by a value of $R_X \ga -3$, and almost all of them belong to a spectral class later than M0. The strongly X-ray-variable sources ($X_{var} \geq 10$) shown in blue in Fig.\ref{fig:Teff_LxLbol} almost all demonstrate $R_X \ga -3$.

\begin{figure*}
  \centering
  
  \includegraphics[width=1.8\columnwidth]{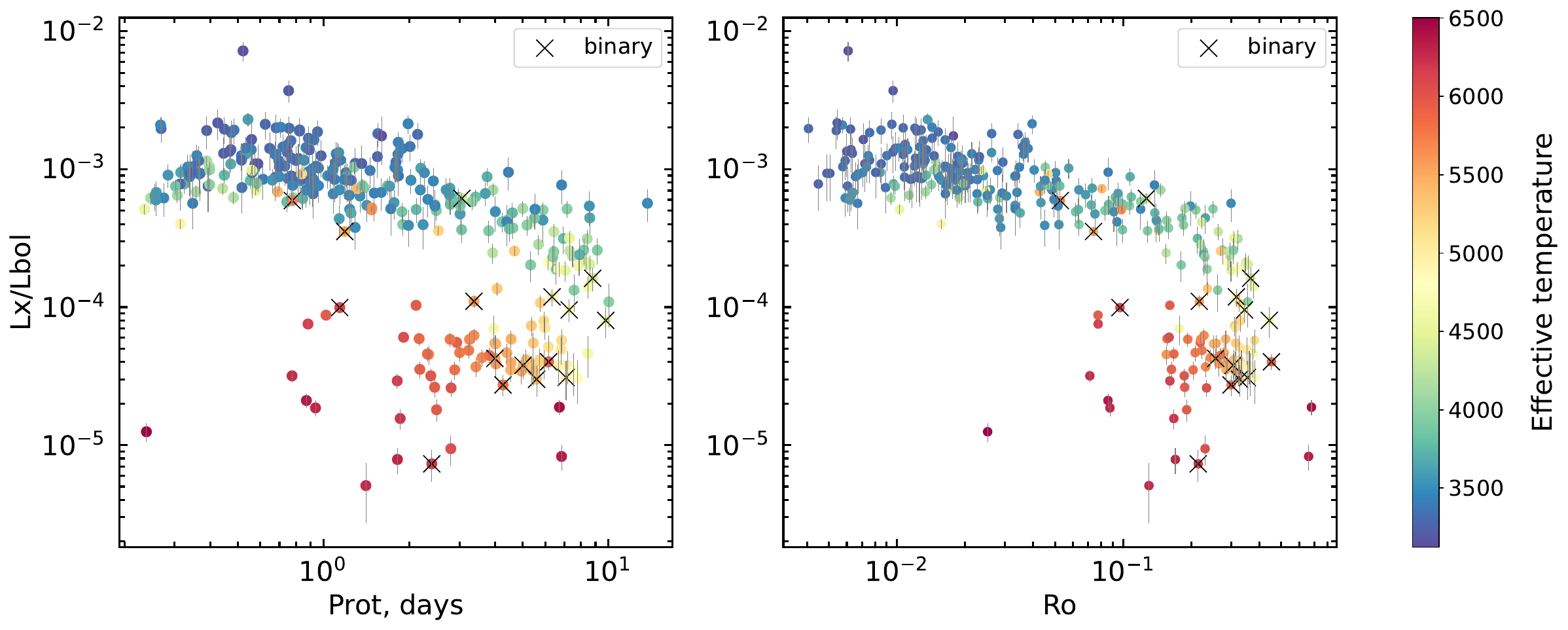}
      
  \hspace{1pt}
  \caption{Dependence of $L_X/L_{bol}$ on the rotation period of the star (from  \cite{2021ApJS..257...46G}) (left panel), and on the Rossby number (right panel).  The spectroscopical binaries \citep{2021ApJ...921..117T} are marked with crosses.}  
  \label{fig:Period_LxLbol}
  
\end{figure*}

\begin{figure*}
  \centering
    \includegraphics[width=1.8\columnwidth]{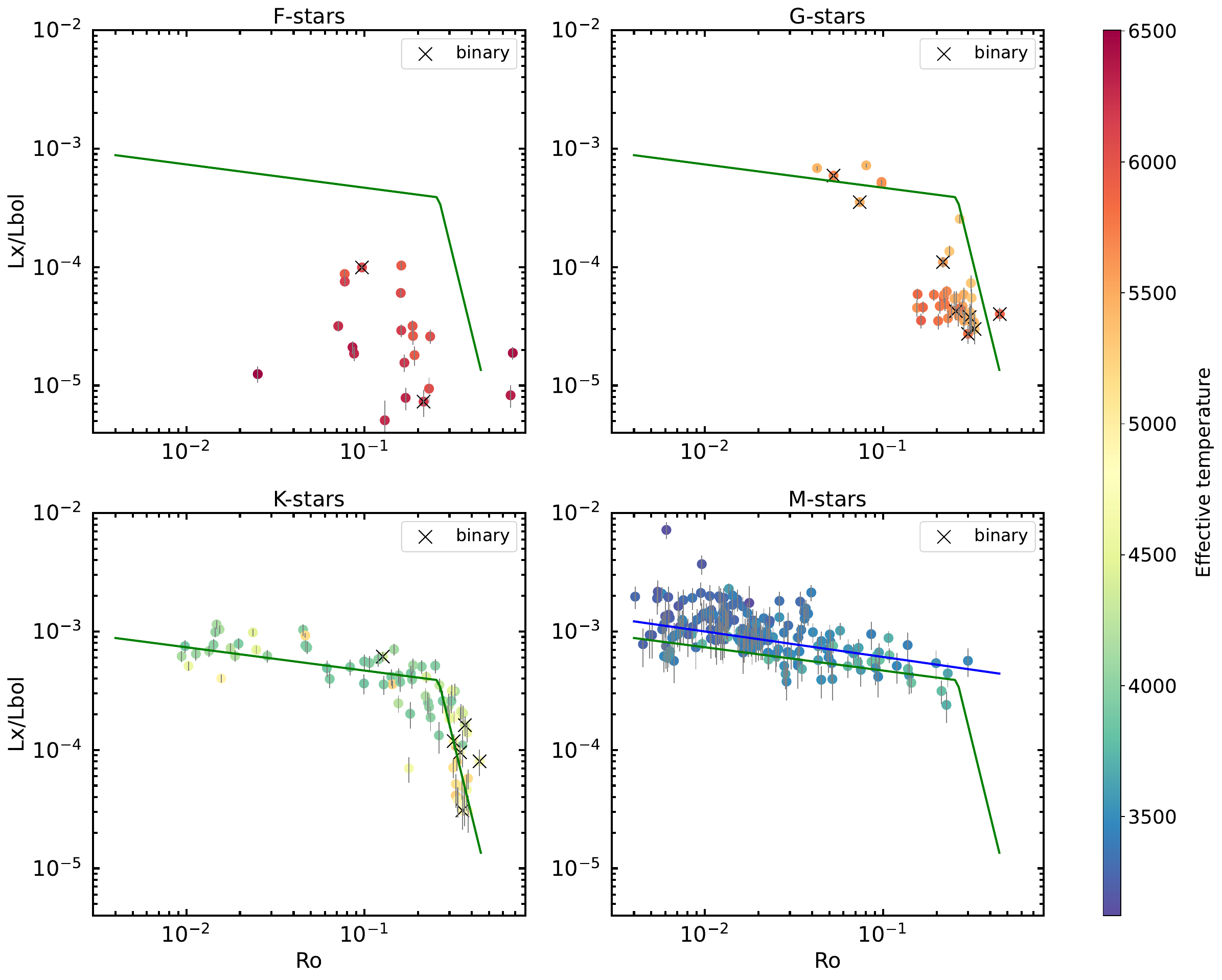}
      
  \hspace{1pt}
  \caption{Dependencies of $L_X/L_{bol}$ on the Rossby number for stars divided by spectral classes F, G, K, and M.  The green line in all plots shows the results of fitting the data for K stars by a power law with a break.  The blue line in the panel for the M stars shows the result of their fitting by the power law. The spectroscopical binaries \citep{2021ApJ...921..117T} are marked with crosses.}
  \label{fig:Ro_LxLbol_FGKM}
  
\end{figure*}

The compiled Pleiades stellar rotation catalog \citep{2021ApJS..257...46G} was used to construct the dependence of $R_X$ on the stellar rotation period (\textit{P}) and a stronger indicator of coronal activity, the Rossby\footnote{https://en.wikipedia.org/wiki/Rossby\_number} number  (\textit{Ro}).
The convective turnover time ($\tau$) required to estimate the Rossby number ($Ro=\frac{P}{\tau}$) was estimated from the empirical dependence on the stellar mass \citep{2011ApJ...743...48W}. This is a $log-log$ second-order polynomial:
\small
{
\begin{equation}
    log(\tau)=1.16-1.49log(M/M_\odot)-0.54log^2(M/M_\odot)
\end{equation}
}
This dependence is valid for the stellar mass range $0.09<M/M_\odot<1.36$, i.e., for stars with a convective zone. In our sample there is a small group of sources with masses higher than $M/M_\odot > 1.36$, but, since they have no convective zone, $\tau$ loses its meaning and they are not used in the construction of dependences on the Rossby number. Sources with $X_{var} \geq 10$ are also excluded from consideration, since the X-ray luminosity of these sources may be overestimated due to observations of them in flares. Sources with more than one Pleiades star from the 6d-catalog in the r98 error circle were also excluded.

The obtained dependences of $L_X/L{bol}$ on the rotation period (left panel) and the Rossby number (right panel) are shown in Fig. \ref{fig:Period_LxLbol}.  Confirmed binary systems are marked with crosses, while sources that have no reliable evidence of multiplicity are marked with different colors which characterize the effective temperature of the star.   

One object in Fig. \ref{fig:Period_LxLbol} -- HD 23912, is located well away from the other stars (lower left corner in the figure). It is a star in which the pecular chemical composition is detected \citep{2021MNRAS.506..150B}. A strong deviation from the general trend as well as a peculiar chemical composition may be in favor of a binary system with a low mass ratio of the components.  

The dependence of $R_X$ on the Rossby number separately for spectral classes F, G, K, and M is shown in Fig.\ref{fig:Ro_LxLbol_FGKM}; for class F, stars later than class F5 were considered. In the dependence for K stars, a slope is clearly visible at the value of the Rossby number $Ro\sim 0.2-0.3$, above which the dependence becomes significantly steeper. This behavior can be qualitatively interpreted as a transition from the unsaturated regime to the saturated regime when the Rossby number decreases below some critical value of $Ro$. For coarse quantitative characterization of the parameters of this dependence, we used approximation by a power law with a break. Since the points have a large spread, much larger than the statistical errors, standard approximation methods such as $\chi^2$ minimization or least squares may lead to results that may depend on the presence of outliers in the sample -- points that deviate significantly from the main trend. Therefore, we applied a simple iterative procedure, somewhat resembling a simplified version of the RANSAC (Random Sample Consensus) algorithm, in which at each iteration a small fraction of $f$ points with the maximum relative deviation in absolute value from the model were excluded. Although this method does not converge in the strict mathematical sense, for small values of $f\sim 1-5\%$, the values of the model parameters stabilize over a wide range of iterations. For $f=1\%$ (which corresponds to the exclusion of one point at each iteration), after 10 iterations we obtained the following parameter values: slope at small and large values of $Ro$ respectively $\beta_1=-0.24\pm0.04$ and $\beta_2=-5.1\pm2.3$, break position $Ro_{break}=0.25\pm0.02$. The approximation was performed in linear coordinates, i.e., for the value $L_X/L_{bol}$ by minimizing the sum of squares of deviations. Minimization of $\chi^2$ leads to similar results.

It is evident from Fig.\ref{fig:Ro_LxLbol_FGKM} that M-class stars (lower right panel) are detected only in the saturation mode. This is due to the lack of measurements of rotation periods for M stars of early classes inhabiting the region of large Rossby numbers. The approximation of the $R_X-Ro$ dependence for M stars by the power law gives a slope of $\beta=-0.23\pm0.02$, close to the corresponding slope for K stars. In the approximation, we used the same approach as for K stars. The size of our sample of stars in the Pleiades does not allow us to characterize the behavior of F- and G-class stars.

Based on a large sample of stars with coronal activity identified using \textit{SRG/eROSITA} data on the western galactic hemisphere, \cite{2024A&A...684A.121F} was determine the position of the break in the $R_X-Ro$ dependence  as $log(Ro,_{sat})=-1.561\pm0.026~dex$. The difference with our results is due to the different determination of the Rossby number and possibly to the fact that \citet{2024A&A...684A.121F} used a broad sample including stars of different ages and chemistries.

\section{Comparison with the \textit{ROSAT} data.}

More than 30 years ago, using the \textit{ROSAT} X-ray telescope and the \textit{PSPC} detector (\cite{1986SPIE..597..208P}, \cite{1991Natur.349..579T}), long-term observations ($\sim~31$Ks) of the core region of the Pleiades cluster with a radius of $50'$ \citep{1996ApJS..102...75M} were made. Coordinates of the field center: $RA=3^h46^m48^s, DEC=23^o54'00''$ [J2000.0]. A total of 99 Pleiades stars with X-rays in the range 0.1--2.1 keV were identified in this field by the \textit{ROSAT}. In the same field, we identified 198 Pleiades stars using the \textit{eROSITA} data.

To search for matches between the \textit{eROSITA} and \textit{ROSAT} sources 
we separately considered three zones of the \textit{ROSAT} field of view as a function of the angular distance to the telescope axis, which determine the positional errors of the \textit{ROSAT} sources. 
Following the approach used in \citet{1996ApJS..102...75M}, matches between the \textit{ROSAT} and \textit{eROSITA} sources were searched in circles with radii of $30''$, $60''$, and $120''$ for the central ($\theta\leq20'$), middle ($20'<\theta\leq30'$), and outer ($30<\theta\leq50'$) regions, respectively. It should be pointed out that the localization accuracy of \textit{eROSITA} sources is much higher, with typical values of $r98$ not exceeding $\sim 10\arcsec$. As a result, 88 matches between the two catalogs were found. And in 22 cases there are 2 sources of \textit{ROSAT} in the error circle.  And for the 11 \textit{ROSAT} sources associated with the Pleiades, X-ray emission is not detected in the \textit{eROSITA} data at the level of sensitivity achieved over 5 surveys.  The \textit{ROSAT} sources with more than one \textit{eROSITA} source in the error circle were not used in the further analysis of variability on a 30-year scale.     

A list of matches between the sources \textit{ROSAT} and \textit{eROSITA} is presented in Table \ref{cat_ROSAT_APJ102}.  

\begin{figure*}
  \centering

\includegraphics[width=1.8\columnwidth,height=0.9\columnwidth]{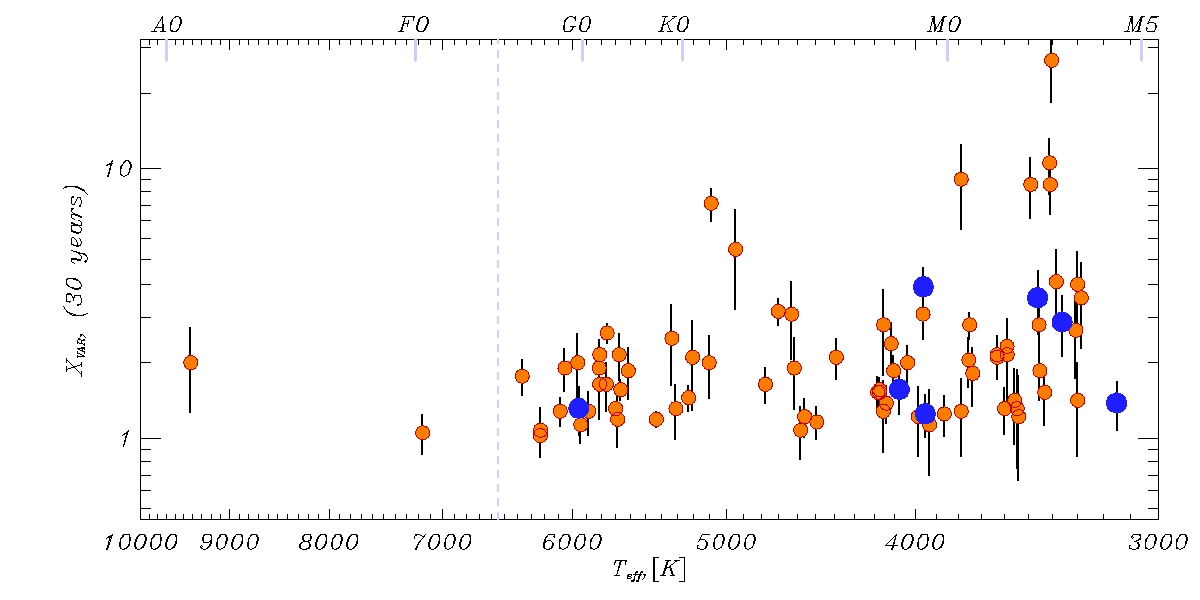}

  \caption{X-ray variability of  Pleiades stars on a 30-year time-scale according to $ROSAT$ and \textit{eROSITA} data as a function of the star's effective temperature $T_{eff}$. The sources whose X-ray variability within the \textit{eROSITA}  5 surveys exceeds a factor of 5 are marked with blue symbols. }
  \label{fig:Teff_eROSITA_ROSAT}
  
\end{figure*}

\begin{figure}
  \centering
  \includegraphics[width=1.0\columnwidth,height=0.85\columnwidth]{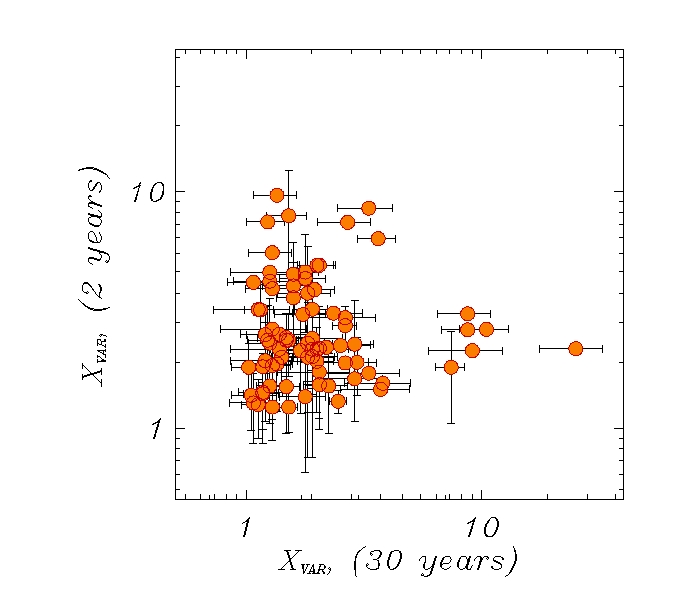}
      
  \hspace{1pt}
  \caption{The relation between the variability factor $X_{var}$ on a time scale of 2 years according to \textit{eROSITA} and on a time scale of 30 years (\textit{ROSAT} and \textit{eROSITA}).}
  \label{fig:Xvar30_02}
  
\end{figure}

A comparison of the \textit{ROSAT} and \textit{eROSITA} catalogs allows us to characterize the variability of stellar X-ray emission on a time scale of $\sim 30$ years. For this purpose, we use the variability factor $X_{var}$.  To compute $X_{var}$, we used only \textit{ROSAT} sources that have only one \textit{eROSITA} source in the error circle.  For each source, using two luminosity values -- $L_{X\_ROSAT}$ and $L_{X\_{eROSITA}}$ -- the factor $X_{var}$ was calculated as the ratio of the larger value to the smaller one.   The dependence of $X_{var}$ on the effective temperature of the star is shown in Fig.\ref{fig:Teff_eROSITA_ROSAT}. The figure shows that several sources of late spectral classes demonstrate variability of the order of or greater than $\sim 10$ times. The other sources are moderately variable, with a flux change factor of up to $\sim 3-4$ times over 30 years. We also note that sources that are highly variable during the \textit{eROSITA} survey -- on the scale of 2 years -- often do not demonstrate strong variability on the scale of 30 years.  This is also confirmed by Fig.\ref{fig:Xvar30_02}, which shows the relationship between $X_{var}$ on a time scale of 2 years according to \textit{eROSITA} and on a time scale of 30 years. It can also be inferred from Fig.\ref{fig:Xvar30_02} that the stars which are highly variable on the 30-yr scale show moderate variability on the 2-yr scale and vice versa. However, larger sample size is required to confirm the validity of this finding and to investigate the reasons for this behavior. This work will be continued in future publications.

\begin{table*}

\caption{List of coincidences of \textit{eROSITA} and \textit{ROSAT} sources in the core region of the Pleiades cluster with radius $ 50' $. The columns are labeled as follows: 1) the source identification number \textit{ROSAT} in Table 1 of the \cite{1996ApJS..102...75M} paper; 2) the source identifier in the \textit{SRG/eROSITA} catalog; 3) the distance between the \textit{ROSAT} and \textit{eROSITA} sources;  4) distance of the X-ray source \textit{ROSAT} from the telescope axis in arcmin; 5) logarithm of the X-ray luminosity of the source according to the \textit{ROSAT} data in the range 0.1-2.1 keV in units of erg/s; 6) logarithm of the X-ray luminosity of the source from the \textit{eROSITA} data in the range 0.3-2.3 keV in units of erg/s; 7) $N_{ero}$ is the number of \textit{eROSITA} sources in the Pleiades that are within the error circle of the \textit{ROSAT} source identified earlier with the cluster star; 8) $N_{gaia}$~- is the number of Pleiades optical sources from the 6d-catalog, based on \textit{Gaia DR3}, in the r98 error circle of the \textit{eROSITA} X-ray source.}   

\label{cat_ROSAT_APJ102}
\centering
\begin{tabular}{lcrrcccc}

\hline
\hline 
$ID$  & SRGe+ & \textbf{$\Delta$}  &  off-axis  & $log(L_X)$  &  $log(L_X)$  &  $N_{ero}$  & $N_{gaia}$ \\
      &       & ($''$)  & ($'$)      & (ROSAT)     &  (eROSITA)   &             &            \\
\hline

  9 & J034342.8+233541 &  23.7 & 46.2 &  29.67 &  29.66 & 1 & 1 \\
 11 & J034350.8+241448 &   7.6 & 45.6 &  29.25 &  29.14 & 2 & 1 \\
 11 & J034351.9+241417 &  41.4 & 45.6 &  29.25 &  28.73 & 2 & 1 \\
 13 & J034351.9+241417 &  92.3 & 43.5 &  29.25 &  28.73 & 2 & 1 \\
 13 & J034356.7+241317 &   9.0 & 43.5 &  29.25 &  28.68 & 2 & 1 \\
 19 & J034411.3+232245 &  15.3 & 47.6 &  29.93 &  29.78 & 1 & 1 \\
 20 & J034412.8+240152 &   7.0 & 36.4 &  29.76 &  29.86 & 1 & 2 \\
 21 & J034412.8+240152 &   4.4 & 36.3 &  29.76 &  29.86 & 1 & 2 \\
 22 & J034414.7+240605 &   7.3 & 37.1 &  29.86 &  30.10 & 1 & 2 \\
 23 & J034416.6+233703 &   3.2 & 38.6 &  29.82 &  28.88 & 1 & 1 \\
 26 & J034423.1+240403 &   4.2 & 34.6 &  29.08 &  29.43 & 1 & 1 \\
 27 & J034423.5+240757 &   4.4 & 35.9 &  29.48 &  29.90 & 1 & 1 \\

 \hline\hline

\end{tabular}
\\

The complete table is available in the archive of the Astronomical Data Center in Strasbourg
\end{table*}

\section{Conclusion.}

The Pleiades, like all open clusters, are being disrupted by the tidal forces imposed by  the Galaxy and giant molecular clouds. In the process, the kinematic parameters of the cluster become strongly distorted.  Most members of young open clusters demonstrate strong magnetic activity, which leads to heating of the coronae of stars up to several million degrees. Thus, the sources are bright in X-rays.  While in isolated solar-type stars the magnetic activity drops on a time scale of $\sim 1$~billion years, it continues for a longer time in cool M-stars due to the interaction of the magnetic field with the stellar wind.  The Pleiades is a fairly young cluster with an age of about 125 million years, therefore it can be expected that the fraction of stars with X-ray emission should be quite high.

The region of sky around the Pleiades open cluster was covered in 5 all-sky surveys carried out by the \textit{eROSITA} X-ray telescope  on board \textit{SRG} space  observatory.   The nominal flux limit of $F_{X,0.3-2.3}\sim7.4\times 10^{-15}~erg/s/cm^2$ in the  0.3--2.3 keV energy range has been achieved. At the distance of the Pleiades (135 pc), it corresponds to an X-ray luminosity $L_{X,0.3-2.3}\sim1.6\times10^{28}$~erg/s. 850 X-ray sources from the \textit{eROSITA} source catalog were associated with stars in the cluster. The Pleiades optical star catalog built on the base of the \textit{GAIA DR3} data was used for the identifications \citep{2023AA...678A..75Z}. In 38 cases, two Pleiades stars were within  98\%  error circle of an X-ray source.  X-ray emission was detected from stars of all spectral types, from  early A-class to the late M-dwarfs (M5V). Taking into account that the typical X-ray luminosity of A-type stars (\cite{2014ApJ...786..136D}, \cite{2022AJ....164....8G}) is several orders of magnitude lower than the luminosity limit reached by \textit{eROSITA}, the X-ray emission of A-class stars is most probably due to the presence of a cold secondary component in the binary system. The \textit{eROSITA} sources associated with the Pleiades emit in X-rays with a total luminosity of $L_{X,total} \sim 1.3 \times 10^{32} $~erg/s. The highest X-ray luminosity is observed for a G-type star and reaches $L_{X,0.3-2.3}\sim2\times10^{30}$~erg/s.

For 27 sources,  \textit{aROSITA}  registered strong X-ray variability between surveys, with a maximum-to-minimum flux ratio of a factor of 10 or more. Most of them are known as eruptive optical variable stars of the dM class.

The logarithm of the ratio of the X-ray to bolometric luminosity  $R_X=log(L_X/L_{bol})$ increases from $-5$ to $-2$ as the effective temperature of the star decreases. The distribution of stars by $R_X$ value is bimodal with maxima around $R_X\sim-4.3$ and $R_X\sim-3.1$.  The left peak with a smaller $R_X$ value is formed by FGK-type stars, while the right peak at $R_X\sim-3.1$ is mainly populated by M-type stars, although other spectral classes with convective envelopes also contribute. The origin of the source locus in the $R_X\sim-4.3$ region is debatable and can be explained either by differences in physical properties between stars of different spectral classes or by SB1-type binaries.

The dependence of $R_X$ on the Rossby number is complex and differs for stars of different spectral classes.  The data for K-type stars can be described by a power law with a break, so that at small Rossby numbers the dependence  of $R_X$ on the Rossby number is weak, and above the threshold value of $Ro\sim 0.25$ it becomes very steep. This behavior can be interpreted as saturation of coronal activity at low Rossby numbers and has already been observed by other authors (\cite{1984A&A...133..117V},\cite{2011ApJ...743...48W},\cite{2024A&A...684A.121F}).  There are no M-type stars with large Rosby number in our sample --  all M-type stars in Pleades show a weak dependence of their $R_X$ on the Rossby number.  In the range of Rossby numbers covered by our sample, $Ro\sim 0.04\div 0.4$, the bulk of G-type stars  have close values of $R_X\sim-4.3$. The distribution of $Ro-R_X$ for F-type stars appears to be more chaotic. We note, however, that our sample size is not enough for a detailed characterization of the behavior of F- and G-type stars.

A comparison with the \textit{ROSAT} catalog was carried out in the region of the cluster core. According to the \textit{eROSITA} data, 198 stars were identified in this region and 88 matches with \textit{ROSAT} were found. For 22 \textit{ROSAT} sources, there are 2 \textit{eROSITA} sources in the error circle.  11 \textit{ROSAT} sources associated with Pleiades stars were not detected by \textit{eROSITA}.  We compared the variability of the sources on time scales of 30 yr and 2 yr.

\acknowledgements
This work is based on observations with the eROSITA
telescope onboard the SRG observatory. The SRG observatory was built by Roskosmos in the interests of the Russian Academy of Sciences represented by its Space Research Institute (IKI) within the framework of the Russian Federal Space Program, with the participation of the Deutsches Zentrum fuer Luft- und Raumfahrt (DLR). The
SRG/eROSITA X-ray telescope was built by a consortium of German Institutes led by MPE, and supported by DLR. The SRG spacecraft was designed, built, launched,
and is operated by the Lavochkin Association and its subcontractors. The science data are downlinked via the
Deep Space Network Antennae in Bear Lakes, Ussurijsk, and Baykonur, funded by Roskosmos. The eROSITA data used in this work were processed using the eSASS
software system developed by the German eROSITA consortium and the proprietary data reduction and analysis software developed by the Russian eROSITA Consortium. This research has made use of the SIMBAD database, operated at CDS, Strasbourg, France.

The authors are grateful to the anonymous reviewer for a thorough and constructive review that helped to improve the paper. The study was supported by a grant from the Russian Science Foundation N  23-12-00292.

\bibliographystyle{astl}
\bibliography{cat.bib}

\begin{thebibliography}{}

\bibitem[\protect\citeauthoryear{{\v{Z}erjal} {et~al.}}{2023}]{2023AA...678A..75Z}
M. {\v{Z}erjal}, N. {Lodieu},  A. {P{\'e}rez-Garrido}, J. {Olivares}, V.~J.~S. {B{\'e}jar}, E.~L. {Mart{\'\i}n}, 
\newblock \aap, {\bf 678}, A75 (2023)

\bibitem[\protect\citeauthoryear{{Lodieu} {et~al.}}{2019}]{2019A&A...628A..66L}
N. {Lodieu}, A. {P{\'e}rez-Garrido}, R.~L. {Smart}, R. {Silvotti}, 
\newblock \aap, {\bf 628}, A66 (2019)

\bibitem[\protect\citeauthoryear{{Vaiana} {et~al.}}{1981}]{1981ApJ...245..163V}
G.~S. {Vaiana}, J.~P. {Cassinelli}, G. {Fabbiano}, L. {Golub}, 
 P. {Gorenstein}, B.~M. {Haisch}, et~al.,
\newblock \apj, {\bf 245}, 163-182 (1981)

\bibitem[\protect\citeauthoryear{{Tr{\"u}mper} {et~al.}}{1991}]{1991Natur.349..579T}
J. {Tr{\"u}mper}, G. {Hasinger}, B. {Aschenbach},  
 H. {Br{\"a}uninger}, U.~G. {Briel},  W. {Burkert},  et~al.,
\newblock Nature, {\bf 349}, 579-583 (1991)

\bibitem[\protect\citeauthoryear{{Giacconi} {et~al.}}{1979}]{1979ApJ...230..540G}
R. {Giacconi}, G. {Branduardi}, U. {Briel},  
 A. {Epstein}, D. {Fabricant}, E. {Feigelson}, et~al.,
\newblock \apj, {\bf 230}, 540-550 (1979)

\bibitem[\protect\citeauthoryear{{Pfeffermann \&} {Briel}}{1986}]{1986SPIE..597..208P}
E. {Pfeffermann}, U.~G. {Briel},  
\newblock in Proceedings, X-ray instrumentation in astronomy, eds. J.~L. Culhane, SPIE,  {\bf 597}, 208-212 (1986)

\bibitem[\protect\citeauthoryear{{Weisskopf} {et~al.}}{2002}]{2002PASP..114....1W}
M.~C. {Weisskopf}, B. {Brinkman}, C. {Canizares}, G. {Garmire}, S.{Murray}, L.~P. {Van Speybroeck}, 
\newblock \pasp, {\bf 114}, 791, 1-24 (2002)

\bibitem[\protect\citeauthoryear{{Jansen} {et~al.}}{2001}]{2001A&A...365L...1J}
F. {Jansen}, D. {Lumb}, B. {Altieri}, J. {Clavel}, M. {Ehle}, C. {Erd}, et~al.,
\newblock \aap, {\bf 365}, L1-L6 (2001)


\bibitem[\protect\citeauthoryear{{Micela} {et~al.}}{1985}]{1985ApJ...292..172M}
G. {Micela}, S. {Sciortino}, S. {Serio}, G.~S. {Vaiana}, J. {Bookbinder},  L. {Golub},  et~al.,
\newblock \apj, {\bf 292}, 172-180 (1985)

\bibitem[\protect\citeauthoryear{{Micela} {et~al.}}{1990}]{1990ApJ...348..557M}
G. {Micela}, S. {Sciortino}, G.~S. {Vaiana}, F.~R., Jr. {Harnden},  R. {Rosner},  J.~H.~M.~M. {Schmitt},
\newblock \apj, {\bf 348}, 557 (1990)


\bibitem[\protect\citeauthoryear{{Schmitt} {et~al.}}{1993}]{1993A&A...277..114S}
J.~H.~M.~M.{Schmitt}, P. {Kahabka}, J. {Stauffer}, A.~J.~M. {Piters}, 
\newblock \aap, {\bf 277}, 114-122 (1993)


\bibitem[\protect\citeauthoryear{{Stauffer} {et~al.}}{1994}]{1994ApJS...91..625S}
J.~R. {Stauffer}, J. -P. {Caillault}, M. {Gagne}, C.~F. {Prosser},  L.~W. {Hartmann}, 
\newblock \apjs, {\bf 91}, 625 (1994)

\bibitem[\protect\citeauthoryear{{Micela} {et~al.}}{1996}]{1996ApJS..102...75M}
G. {Micela}, S. {Sciortino}, V. {Kashyap}, F.~R., Jr. {Harnden}, R. {Rosner}, 
\newblock \apjs, {\bf 102}, 75 (1996)

\bibitem[\protect\citeauthoryear{{Sunyaev} {et~al.}}{2021}]{2021AA...656A.132S}
R. {Sunyaev}, V. {Arefiev}, V. {Babyshkin}, A. {Bogomolov}, K. {Borisov}, M.
  {Buntov}, et~al.,
\newblock \aap, {\bf 656}, A132, 29 (2021)

\bibitem[\protect\citeauthoryear{{Predehl} {et~al.}}{2021}]{2021AA...647A...1P}
P. {Predehl}, R. {Andritschke}, V. {Arefiev}, V. {Babyshkin}, O. {Batanov}, W.
  {Becker}, et~al.,
\newblock \aap, {\bf 647}, A1, 16 (2021)

\bibitem[\protect\citeauthoryear{{Krishnamurthi} {et~al.}}{2001}]{2001AJ....121..337K}
Anita {Krishnamurthi}, Christopher S. {Reynolds}, Jeffrey L. {Linsky}, Eduardo {Mart{\'\i}n}, Marc {Gagn{\'e}}, 
\newblock \aj, {\bf 121}, 1, 337-346 (2001)

\bibitem[\protect\citeauthoryear{{Daniel} {et~al.}}{2002}]{2002ApJ...578..486D}
Kathryne J. {Daniel}, Jeffrey L. {Linsky}, Marc {Gagn{\'e}}, 
\newblock \apj, {\bf 578}, 1, 486-502 (2002)

\bibitem[\protect\citeauthoryear{{Briggs \&} {Pye}}{2003}]{2003MNRAS.345..714B}
K.~R. {Briggs}, J.~P. {Pye}, 
\newblock \mnras, {\bf 345}, 3, 714-726 (2003)

\bibitem[\protect\citeauthoryear{{Guarcello} {et~al.}}{2019}]{2019A&A...622A.210G}
M.~G. {Guarcello}, G. {Micela}, S. {Sciortino}, J. {L{\'o}pez-Santiago}, C. {Argiroffi}, F. {Reale}, et~al., 
\newblock \aap, {\bf 622}, A210 (2019)

\bibitem[\protect\citeauthoryear{{Pecaut \&} {Mamajek}}{2013}]{2013ApJS..208....9P}
Mark J. {Pecaut}, Eric E. {Mamajek}, 
\newblock \apjs, {\bf 208}, 1, 9 (2013)

\bibitem[\protect\citeauthoryear{{Bressan} {et~al.}}{2012}]{2012MNRAS.427..127B}
Alessandro {Bressan}, Paola {Marigo}, L{\'e}o. {Girardi}, Bernardo {Salasnich}, Claudia {Dal Cero}, Stefano {Rubele}, Ambra {Nanni}, 
\newblock \mnras, {\bf 427}, 1, 127-145 (2012)

\bibitem[\protect\citeauthoryear{{Netopil} {et~al.}}{2016}]{2016A&A...585A.150N}
M. {Netopil}, E. {Paunzen},  U. {Heiter}, C. {Soubiran}, 
\newblock \aap, {\bf 585}, A150 (2016)

\bibitem[\protect\citeauthoryear{{Bell} {et~al.}}{2012}]{2012MNRAS.424.3178B}
Cameron P.~M. {Bell},  Tim {Naylor},  N.~J. {Mayne}, R.~D. {Jeffries},  S.~P. {Littlefair}, 
\newblock \mnras, {\bf 424}, 4, 3178-3191 (2012)

\bibitem[\protect\citeauthoryear{{Somers} {et~al.}}{2020}]{2020ApJ...891...29S}
Garrett {Somers},  Lyra {Cao},  Marc H. {Pinsonneault},  
\newblock \apj, {\bf 891}, 1, 29 (2020)

\bibitem[\protect\citeauthoryear{{Drake} {et~al.}}{2014}]{2014ApJ...786..136D}
Jeremy J. {Drake}, Jonathan {Braithwaite}, Vinay {Kashyap}, H. Moritz {G{\"u}nther},  Nicholas J. {Wright}, 
\newblock \apj, {\bf 786}, 2, 136 (2014)

\bibitem[\protect\citeauthoryear{{G{\"u}nther} {et~al.}}{2022}]{2022AJ....164....8G}
Hans Moritz {G{\"u}nther}, Carl {Melis}, J. {Robrade}, P.~C. {Schneider}, Scott J. {Wolk}, Rakesh K. {Yadav}, 
\newblock \aj, {\bf 164}, 1, 8 (2022)

\bibitem[\protect\citeauthoryear{Vilhu}{1984}]{1984A&A...133..117V}
O. {Vilhu},
\newblock \aap, {\bf 133}, 117-126 (1984)

\bibitem[\protect\citeauthoryear{{Wright} {et~al.}}{2011}]{2011ApJ...743...48W}
Nicholas J. {Wright}, Jeremy J. {Drake}, Eric E. {Mamajek}, Gregory W. {Henry}, 
\newblock \apj, {\bf 743}, 1, 48 (2011)

\bibitem[\protect\citeauthoryear{{Freund} {et~al.}}{2024}]{2024A&A...684A.121F}
S. {Freund}, S. {Czesla},  P. {Predehl}, J. {Robrade}, M. {Salvato}, P.~C. {Schneider},  et~al.
\newblock \aap, {\bf 684}, A121 (2024)

\bibitem[\protect\citeauthoryear{{G{\"u}del \&} {Naz{\'e}}}{2009}]{2009A&ARv..17..309G}
Manuel {G{\"u}del}, Ya{\"e}l {Naz{\'e}},
\newblock Astron. and Astroph. Rev., {\bf 17}, 3, 309-408 (2009)


\bibitem[\protect\citeauthoryear{{N{\'u}{\~n}ez} {et~al.}}{2022}]{2022ApJ...931...45N}
Alejandro {N{\'u}{\~n}ez}, Marcel A. {Ag{\"u}eros}, Kevin R. {Covey}, Stephanie T. {Douglas}, Jeremy J. {Drake}, Rayna {Rampalli},  et~al.
\newblock \apj, {\bf 931}, 1, 45 (2022)


\bibitem[\protect\citeauthoryear{{Godoy-Rivera} {et~al.}}{2021}]{2021ApJS..257...46G}
Diego {Godoy-Rivera}, Marc H. {Pinsonneault}, Luisa M. {Rebull}, 
\newblock \apjs, {\bf 257}, 2, 46 (2021)


\bibitem[\protect\citeauthoryear{{Torres} {et~al.}}{2021}]{2021ApJ...921..117T}
Guillermo {Torres}, David W. {Latham}, Samuel N. {Quinn}, 
\newblock \apj, {\bf 921}, 2, 117 (2021)


\bibitem[\protect\citeauthoryear{{Buder} {et~al.}}{2021}]{2021MNRAS.506..150B}
Sven {Buder}, Sanjib {Sharma}, Janez {Kos},  Anish M. {Amarsi}, Thomas {Nordlander}, Karin {Lind},  et~al.  
\newblock \mnras, {\bf 506}, 1, 150-201 (2021)




\end{thebibliography}

\end{document}